\documentclass[aps,prd,preprint,groupedaddress]{revtex4-1}

\usepackage{amssymb,amsmath}
\usepackage{dcolumn}
\usepackage{bm}

\renewcommand{\Re}{ {\mathbf{Re} }}

\renewcommand{\det}{ {\mathrm{det} }}

\begin{document}


\title{Analytic QCD Binding Potentials}


\author{H. M. Fried$^{\star}$, Y. Gabellini$^\dagger$, T. Grandou$^\dagger$, Y.-M. Sheu$^{\dagger}$}
\affiliation{${}^\star$ {Physics Department, Brown University, Providence, RI 02912, USA} \\ ${}^\dagger$ {Universit\'{e} de Nice Sophia-Antipolis, Institut Non Lin$\acute{e}$aire de Nice, UMR 6618 CNRS, 06560 Valbonne, France}}


\date{\today}

\begin{abstract}
This paper applies the analytic forms of a recent non-perturbative, manifestly gauge- and Lorentz-invariant description (of the exchange of all possible virtual gluons between quarks ($Q$) and/or anti-quarks ($\bar{Q}$) in a quenched, eikonal approximation) to extract analytic forms for the binding potentials generating a model $Q$-$\bar{Q}$ "pion" , and a model $QQQ$ "nucleon". Other, more complicated $Q$, $\bar{Q}$ contributions to such color-singlet states may also be identified analytically.

An elementary minimization technique, relevant to the ground states of such bound systems, is adopted to approximate the solutions to a more proper, but far more complicated Schroedinger/Dirac equation; the existence of possible contributions to the pion and nucleon masses due to spin, angular momentum, and "deformation" degrees of freedom is noted but not pursued. Neglecting electromagnetic and weak interactions, this analysis illustrates how the one new parameter making its appearance in this exact, realistic formalism may be evaluated, along with a qualitative estimate of the lowest quark mass and the energy contributions to the ground states of the gluon fields, from a knowledge of the pion and (ground state) nucleon masses.

\end{abstract}


\maketitle


\section{\label{SEC:Intro}Introduction}

This paper should be viewed as a continuation of that of Ref.~\cite{Fried2009}, in which a new formalism for analytic QCD calculations was suggested, and which can serve to identify an effective potential as the result of the exchange of multiple, virtual gluons between $Q$'s and/or $\bar{Q}$'s.  The word 'effective' here represents the type of potential one would expect to find were the scattering pair of conventional Abelian particles, moving at large relative velocities in their CM, as appropriate in an eikonal context.  And the word 'Abelian' is most relevant in this QCD situation, for the analysis of Ref.~\cite{Fried2009} clearly shows that for large impact parameters the scattering of such non-Abelian $Q$s and/or $\bar{Q}$s is effectively coherent; only as the impact parameter decreases do progressively stronger color fluctuations appear and decrease the the effective coupling, which then becomes the statement of non-perturbative asymptotic freedom in the variable conjugate to impact parameter.

In the present paper, we shall be interested in binding, or "restoring" potentials, effective potentials which prevent bound particles from remaining at relatively large separations.  Since large impact parameters are relevant to large, three-dimensional distances, what is needed is, firstly, a statement of the relevant formulae of Ref.\cite{Fried2009} at large impact parameter; and, secondly, the corresponding eikonal function found in that limit.  It is then a relatively simple matter to determine an effective potential which would produce the same eikonal.

There is one truly profound statement which deserves careful mention, as in Ref.~\cite{Fried2009} and in the paper immediately preceding this one~\cite{Fried2011_QCD3}, which is that the quanta of the the non-Abelian QCD fields differ from quanta of Abelian fields in that the latter quanta, in principle, satisfy standard quantum-mechanical measurement possibilities: position can be specified exactly at the expense of momenta, and conversely. Quarks and gluons, in contrast, are bound objects; and in the context of $Q$/$\bar{Q}$ scattering, while their longitudinal and time coordinates (or their longitudinal momentum and energy) can be estimated in terms of that of the hadron into which they are bound, their transverse momenta, or the (impact parameter) distance between them, can never be measured with precision.  It is precisely here that one is forced to bridge the gap from "ideal" to "realistic" QCD; and until a more fundamental field theory of QCD is developed, which contains such imprecision from its very beginning, one will be forced to make a "phenomenological" transition into the "realistic" world of physical QCD, as in Ref.~\cite{Fried2011_QCD3}.

The need for such a transition was made quite clear in Ref.~\cite{Fried2009}, where one of a variety of forms was suggested as appropriate, when it became clear that a $\delta^{(2)}(\vec{b})$ must be replaced by a smoothly-falling function of impact parameter, \emph{e.g.}, $\varphi(b) \sim e^{-(\mu b)^{2}}$, although no definite choice was needed to reach the conclusions of that paper.  In the present context of binding, a specific choice is made; but it should be emphasized that this choice is phenomenological, if reasonable: it works, generating a binding potential $V(r)$ proportional to $\xi \cdot r^{1+\xi}$, where $\xi$ is a small, real and positive parameter.  The appearance of $\xi$ was unexpected, and represents a new quantity which should appear naturally in a more profound version of QCD; it is here specified by a minimization procedure of the ground-state's energy level of our model pion.

It should also be emphasized that the transition to "realistic" QCD has nothing to do with the quenched, eikonal approximations used to give the simplest possible, non-trivial example of our new approach to gauge-invariant, non-perturbative QCD.  Exactly the same necessity of such a transition may be seen in the second paper of this series~\cite{Fried2010_QCD2}, where the forms of the results displayed in Ref.~\cite{Fried2009} are showed to hold without approximation and without exception.

Our present discussion will follow that of Ref.~\cite{Fried2009} in extracting, under quenched approximation, the eikonal and a corresponding potential function for the equal-mass $Q$-$\bar{Q}$ contribution to the model singlet state we here call the "pion".  To make a similar statement for the $QQQ$ contribution to the model singlet ground state we call the "nucleon", it will be useful to recall the eikonal forms appropriate to three-body scattering amplitudes~\cite{Ref4,HMF2}, while for application to higher $Q$/$\bar{Q}$ contributions to these and other hadrons, the discussion of N-body eikonals of Ref.~\cite{Ref4,Fried1983} will suffice.

We refer the reader to the unique method employed in Refs.~\cite{Fried2009,Fried2010_QCD2} to convert Schwinger's conventional, gauge-dependent functional representation for the QCD generating functional into an exact, related representation, in which all processes corresponding to $Q$ and $\bar{Q}$ interactions by the (non-perturbative) exchange of all possible gluons -- including  $Q$ and $\bar{Q}$, cubic and quartic interactions -- are rigorously and manifestly gauge invariant.  Attention is then drawn to the quenched $Q$-$\bar{Q}$ scattering amplitude in eikonal approximation, and to the manipulations and (strong-coupling) simplification used to obtain the integral of Eq.~(44) of Ref.~\cite{Fried2009}, and the replacement of its integrand by Eq.~(60).  We begin our extraction of the $Q$-$\bar{Q}$ binding potential from this point.


\section{\label{SEC:SEC2}Quark Binding Potential}

In this Section, we shall be interested in binding, or "restoring" potentials, effective potentials which prevent bound particles from remaining at relatively large separations.  Since large impact parameters are relevant to large, three-dimensional distances, what is needed is, firstly, a statement of the relevant formulae of Ref.~\cite{Fried2009} at large impact parameter; and, secondly, the corresponding eikonal function found in that limit.  It is then a relatively simple matter to determine an effective potential which would produce the same eikonal.

A two-body eikonal scattering amplitude will have the form~\cite{Ref4}
\begin{eqnarray}\label{Eq:1}
\mathcal{T}(s,t) = \frac{is}{2m^{2}} \int{\mathrm{d}^{2}\vec{b} \, e^{i \vec{q} \cdot \vec{b}} \left[ 1 - e^{i\mathbf{X}(s,\vec{b})}\right]},
\end{eqnarray}

\noindent where $s$ and $t$ denote the standard Mandelstam variables, $s = -(p_{1} + p_{2})^{2}$, $t = -(p_{1} - p'_{1})^{2} = \vec{q}^{\, 2}$ in the CM of $Q\bar{Q}$, and $\mathbf{X}(s,b)$ is the eikonal function.  Following the arguments of Ref.~\cite{Fried2009}, one reaches
\begin{eqnarray}\label{Eq:2}
e^{i\mathbf{X}} = \mathcal{N} \int{\mathrm{d}^{n} \bar{\chi}_{30} \, \left[ \det(g f \cdot \bar{\chi})^{-1}\right]^{\frac{1}{2}} \cdot e^{i \frac{\bar{\chi}^{2}}{4}} \cdot \exp{\left[ i g \varphi(b) \, \Omega^{a}_{\mathrm{I}} \cdot \left. (f \cdot \bar{\chi})^{-1}\right|^{ab}_{30} \cdot \Omega^{a}_{\mathrm{I\!I}} \right]  }}
\end{eqnarray}

\noindent and the normalization is such that for $\varphi(b) \rightarrow 0$, $\exp{[i \mathbf{X}]} \rightarrow 1$.  All the integrals over
\begin{eqnarray}
\int{\mathrm{d}^{n} \alpha_{\mathrm{I}}} \, \int{\mathrm{d}^{n} \alpha_{\mathrm{I\!I}}} \, \int{\mathrm{d}^{n} \Omega_{\mathrm{I}}} \, \int{\mathrm{d}^{n} \Omega_{\mathrm{I\!I}}}
\end{eqnarray}

\noindent are properly normalized, and they connect the $\Omega^{a}_{\mathrm{I}}$, $\Omega^{b}_{\mathrm{I\!I}}$ dependence of Eq.~(\ref{Eq:2}) with the $\lambda^{a}_{\mathrm{I}}$, $\lambda^{b}_{\mathrm{I\!I}}$ Gell-Mann matrices placed between initial and final state vectors, which define the type of amplitude desired, $Q$-$Q$ scattering or $Q$-$\bar{Q}$ scattering, \emph{etc.}

As in Ref.~\cite{Fried2009}, we arrange the 8-dimensional integral over $\int{\mathrm{d}^{n} \bar{\chi}_{03}}$ into the form of a radial integration over the magnitude of $\bar{\chi}_{03}$, now called $R$, and normalized angular integrations.  We suppress the latter, along with all normalized $ \int{\mathrm{d}^{n} \Omega_{\mathrm{I}}} \, \int{\mathrm{d}^{n} \Omega_{\mathrm{I\!I}}}$ dependence, and rewrite Eq.~(\ref{Eq:2}) in the form
\begin{eqnarray}\label{Eq:3}
e^{i\mathbf{X}(s,b)} = \mathcal{N}' \int_{0}^{\infty}{\mathrm{d}R \, R^{6.5} \, e^{i \frac{R^{2}}{4} + i \langle g\rangle \frac{\varphi(b)}{R} } }
\end{eqnarray}

\noindent where the normalization of the $R$-integral is given by ${\mathcal{N}'}^{-1} = \int_{0}^{\infty}{\mathrm{d}R \, R^{6.5} \, e^{i \frac{R^{2}}{4} }}$, and where $\langle g \rangle$ is a shorthand denoting the effective coupling AFTER all angular and color integrations and matrix elements have been calculated.  In this calculation, only the linear dependence of $g \varphi$ is needed, while the non-vanishing matrix elements between singlet color states will define $Q$-$\bar{Q}$ scattering.

As noted in Section \ref{SEC:Intro}, we wish to compare this eikonal function at fairly large impact parameter with that corresponding to the scattering of two, interacting Abelian particles, and then infer what form of effective potential corresponds to that eikonal. We therefore assume the usual formula~\cite{Ref4} relating an eikonal to a specified potential,
\begin{eqnarray}\label{Eq:4}
\mathbf{X}(s,b) = - \int_{-\infty}^{\infty}{\mathrm{d}z \, V(\vec{b}+\hat{\mathrm{P}}_{\mathrm{L}}z)} = \gamma(s) \, \mathbf{X}(b),
\end{eqnarray}

\noindent where $\gamma(s)$ depends upon the CM energy of the scattering particles (and the nature of their interaction), and $\hat{\mathrm{P}}_{\mathrm{L}}$ is a unit vector in the direction of longitudinal motion.  Eq.~(\ref{Eq:4}) is true for non-relativistic and relativistic scattering, in potential theory and in field theory; for large $b$, our eikonal will turn out to be only weakly dependent on $s$, and the $b$-independent contributions do not contribute to $V(r)$.  To obtain $V(r)$ from a given  $\mathbf{X}(b)$ is then simply a matter of reversing the usual calculation: one computes the Fourier transform $\tilde{\mathbf{X}}(k_{\perp})$ of  $\mathbf{X}(b)$, and then extends $k_{\perp}$ to $|\vec{k}| = [ k_{\perp}^{2} + k_{\mathrm{L}}^{2} ]^{\frac{1}{2}}$.  The three-dimensional Fourier transform of $\tilde{\mathbf{X}}(k)$ is then proportional to $V(r)$.  Following the discussion of Ref.~\cite{Fried2011_QCD3}, we assume that $\mathbf{X}(b)$ -- which depends on the assumed form of $\varphi(b)$ -- is a function of $b^{2}$; and although we begin by asking for the form of $\varphi(b)$ for large $b$, it can be shown that the small-$b$ contributions to the Fourier transforms are unimportant for the large-$r$ behavior of $V(r)$.

There is one further point which requires consideration, for the above remarks are valid when the process involves only the scattering of two particles.  But when initial energies are high enough such that inelastic particle production occurs, then the potential corresponding to such production must have a negative imaginary component (so that the corresponding S-matrix, $ \exp{[-i\mathcal{H}t]}$, will have a decaying time dependence), thereby conserving probability as a diminution of the final, two-particle state when production becomes possible.  And this is also true when the two initial particles have the possibility of binding, and forming a state which was not initially present; the probability of the final scattering state must diminish.  In other words, the eikonal function one expects to find may always be characterized by a complex potential of form $V_{\mathrm{R}} - i V_{\mathrm{I}}$, where in the present case, $V_{\mathrm{I}}$ is that potential which can bind the $Q$ and $\bar{Q}$ into a pion; and that potential, in a scattering calculation, must appear with a multiplicative factor of $- i$.  Symbolically, $i\mathbf{X} \rightarrow i(V_{\mathrm{R}} - i V_{\mathrm{I}})$, and for this binding calculation, $i\mathbf{X} \rightarrow V_{\mathrm{B}}$.

Return to Eq.~(\ref{Eq:1}) and assume $\vec{q} \neq 0$, so that the "1" term of the integrand contributes a $\delta^{(2)}(q)$ which vanishes.  This is a physical assumption: $Q$ and $\bar{Q}$ are not bound rigidly, with a unchanging impact parameter; rather, there is a constant, if small, "in-and-out" transverse motion, which represents the bound state; and hence $\vec{q} \neq 0$.  Then,
\begin{eqnarray}
\mathcal{T} \sim  \int{\mathrm{d}^{2}\vec{b} \, e^{i \vec{q} \cdot \vec{b}} e^{i\mathbf{X}(b)} }
\end{eqnarray}

\noindent where $\exp{[i \mathrm{X}(b)]}$ is given by the normalized Eq.~(\ref{Eq:3}).  In Ref.~\cite{Fried2009} we imagined that $\varphi(b)$ would be a smooth function which vanishes for large $b$, such as $\sim \exp{[-(\mu b)^{2}]}$, but it turns out, below, that an appropriate choice is
\begin{eqnarray}
\varphi(b) = \varphi(0) \, e^{- (\mu b)^{2+\xi} },
\end{eqnarray}

\noindent where $\xi$ is real, positive, and small.  In the preceding paper, Ref.~\cite{Fried2011_QCD3}, a detailed argument for such a "realistic" choice of $\varphi(b)$ was given, leading to the generic Eq.~(21) of that paper.  For any such choice of positive $\xi$, $\varphi$ becomes small when $\mu b \gg 1$.

For large $b$ it is then sensible to expand the $\exp{[i \langle g \rangle \varphi(b) / R]}$ term of Eq.~(\ref{Eq:3}), retaining only the linear $\langle g \rangle \varphi$ dependence.  This gives, in place of Eq.~(\ref{Eq:3}),
\begin{eqnarray}\label{Eq:5}
e^{i\mathbf{X}(b)} &\simeq& \mathcal{N}' \int_{0}^{\infty}{\mathrm{d}R \, R^{6.5} \, e^{i \frac{R^{2}}{4}} \left[1 + i \langle g\rangle \frac{\varphi(b)}{R} + \cdots \right] } \\ \nonumber &=& 1 + i \kappa \mathcal{N}' \langle g\rangle \varphi(b) + \cdots,
\end{eqnarray}

\noindent where
\begin{eqnarray}
\kappa \mathcal{N}' = \mathcal{N}' \int_{0}^{\infty}{\mathrm{d}R \, R^{5.5} \, e^{i \frac{R^{2}}{4}} } \\ \nonumber = \frac{1}{2} (i)^{-\frac{1}{2}} \frac{\Gamma(\frac{13}{4})}{\Gamma(\frac{15}{4})},
\end{eqnarray}

\noindent and $\mathcal{N}' = 2 (4i)^{-\frac{15}{4}} / {\Gamma(\frac{15}{4})} $.  Remembering that both sides of Eq.~(\ref{Eq:5}) are to be integrated over $\int{\mathrm{d}^{2}\vec{b} \, e^{i \vec{q} \cdot \vec{b}} }$, for $\vec{q} \neq 0$ we can associate
\begin{equation}
e^{i\mathbf{X}(b)} = i \kappa \mathcal{N}' {\langle g \rangle} \varphi(b) ,
\end{equation}

\noindent or
\begin{equation}
i\mathbf{X}(b) = \ln{\left[ \varphi(b)\right]} + \cdots,
\end{equation}

\noindent where $\ln{\left[ \varphi \right]}$ is large and $\varphi$ is (effectively) small.  Choosing $\varphi(b) = \varphi(0) \, e^{- (\mu b)^{2+\xi}}$, we then obtain
\begin{equation}\label{Eq:6}
i\mathbf{X}(b) = - (\mu b)^{2+\xi} + \cdots
\end{equation}

It is then convenient to use the integrals~\cite{GR1965}
\begin{equation}
\int_{0}^{\infty}{\mathrm{d}x \, x^{\mu} \, J_{0}(ax)} = 2^{\mu} \, a^{-1-\mu} \, {\Gamma(\frac{1}{2} + \frac{\mu}{2})} / {\Gamma(\frac{1}{2} - \frac{\mu}{2})}, \quad \mu < \frac{1}{2},
\end{equation}

\noindent and
\begin{equation}
\int_{0}^{\infty}{\mathrm{d}x \, x^{\mu - 1} \, \sin(x)} = \Gamma(\mu) \, \sin(\frac{\mu \pi}{2}), \quad |\Re{(\mu)}| < 1,
\end{equation}

\noindent in which, except for obvious poles, the Gamma functions are analytic in $\mu$, and can be continued to the needed values.  With
\begin{eqnarray}
i\tilde{\mathbf{X}}(k_{\perp}) &=& - \int{\mathrm{d}^{2}b \, (\mu b)^{2+\xi} \, e^{i \vec{k}_{\perp} \cdot \vec{b}} } \\ \nonumber &=& - (2 \pi)\int_{0}^{\infty}{\mathrm{d}b \, J_{0}(k_{\perp} b) \, \mu^{2+\xi} \, b^{3+\xi} },
\end{eqnarray}

\noindent and the use of the doubling formula for Gamma functions, working everything through, one finds
\begin{equation}
V_{\mathrm{B}}(r) = -\frac{2^{3+\xi}}{\pi} \, \mu^{2+\xi} \, r^{1+\xi} \, \frac{\Gamma(2 + \frac{\xi}{2})}{\Gamma(-1 - \frac{\xi}{2})} \, \Gamma(-2-\xi) \, \sin(\frac{\pi  \xi}{2}),
\end{equation}

\noindent and for small $\xi$, $\xi \ll 1$,
\begin{equation}\label{Eq:7}
V_{\mathrm{B}}(r) \simeq \xi \, \mu \, (\mu r)^{1+\xi},
\end{equation}

\noindent which may be compared to the results of several machine groups.
There remains the question of what values should be assigned to $\xi$, and to the unknown parameter $\mu / m_{\mathrm{Q}}$; and these will be dealt with in the next Section.

To calculate the corresponding, effective, restoring potential when one of the three quarks contributing to the $QQQ$ bound state is separated from the other two, or, more simply, when all three quarks are forced to separate from each other, one may refer to Eq.~(22) of Ref.~\cite{Fried1983}, giving the eikonal amplitude corresponding to Coulomb three-particle scattering.  Here, the specifically Coulomb parts of this amplitude may be replaced by the single Halpern integral which connects the three quarks to each other, as it connected the $Q$ and $\bar{Q}$ to each other in the calculation above.  If particle 2 of this formula enters the scattering with zero transverse momentum in the rest frame of particles 1 and 3, the amplitude simplifies to
\begin{eqnarray}\label{Eq:8}
\mathcal{T}_{3}^{\mathrm{eik}} \sim \int{\mathrm{d}^{2}\vec{b}_{12} \, \int{\mathrm{d}^{2}\vec{b}_{32} \, e^{i \vec{q}_{3} \cdot \vec{b}_{32} + i \vec{q}_{1} \cdot \vec{b}_{12} } \left[ \Phi(\vec{b}_{12}, \vec{b}_{32}, \vec{b}_{13}) + \Psi_{3} \right] }},
\end{eqnarray}

\noindent where $\vec{b}_{13} = \vec{b}_{12} - \vec{b}_{32}$, irrelevant kinematic factors multiplying these integrals have been suppressed, and $\Psi_{3}$ denotes the combination
\begin{equation}
-1 + \left[ 1 - \Phi(\vec{b}_{12}, \infty, \infty) \right] + \left[ 1 - \Phi(\infty, \vec{b}_{32}, \infty) \right] + \left[ 1 - \Phi(\infty, \infty, \vec{b}_{13}) \right],
\end{equation}

\noindent so that $\mathcal{T}_{3}^{\mathrm{eik}}$ is a completely "connected" amplitude.

Writing $\Phi$ as $\exp{[i \mathbf{X}(b_{12}, b_{32})]}$, its defining integral may be rewritten as
\begin{equation}\label{Eq:9}
\mathcal{N}' \int_{0}^{\infty}{\mathrm{d}R \, R^{6.5} \, \exp{\left[ i \frac{R^{2}}{4} + i \left[ {\langle g \rangle}_{12} \varphi(b_{12}) + {\langle g \rangle}_{32} \varphi(b_{32}) + {\langle g \rangle}_{13} \varphi(b_{13}) \right]\right]} },
\end{equation}

\noindent and one may ask for the form Eq.~(\ref{Eq:9}) takes when one or more of the $b_{ij}$ becomes large.

When any one of the $b_{ij}$ becomes large, its $\varphi(b_{ij})$ becomes small, and its exponential term may be expanded.  The simplest situation is when they all become large, so that an expansion of Eq.~(\ref{Eq:9}) is relevant.  The first non-zero and leading term in such an expansion which contains the $\varphi$ dependence, and hence the $b_{ij}$-dependence, of all three quarks contributing to the singlet $QQQ$ state is that term in which the expansion yields
\begin{equation}\label{Eq:10}
1 + \kappa \mathcal{N}' \left[ {\langle g \rangle}_{12} \varphi(b_{12}) + {\langle g \rangle}_{32} \varphi(b_{32}) + {\langle g \rangle}_{13} \varphi(b_{13}) \right] + \cdots,
\end{equation}

\noindent so that, suppressing all normalized angular and color integrations, for $\vec{q}_{1} \neq 0$, $\vec{q}_{3} \neq 0$,
\begin{equation}\label{Eq:11}
e^{i \mathbf{X}(b_{12}, b_{32}) } \simeq \kappa \mathcal{N}' \left[ {\langle g \rangle}_{12} \varphi(b_{12}) + {\langle g \rangle}_{32} \varphi(b_{32}) + {\langle g \rangle}_{13} \varphi(b_{13}) \right] + \cdots.
\end{equation}

\noindent If particles 1 and 3 are now additionally separated, keeping the distances $b_{12}$ and $b_{32}$ essentially fixed, then
\begin{eqnarray}\label{Eq:12}
e^{i \mathbf{X}(b_{12}, b_{32}) } &\simeq& \kappa \mathcal{N}' {\langle g \rangle}_{13} \varphi(b_{13}) + \cdots, \\ \nonumber i \mathbf{X}(b_{12}, b_{32}) &\simeq& \ln{\left[\varphi(b_{13})\right]} + \cdots,
\end{eqnarray}

\noindent and we are effectively back in the "pion" situation, where the large-impact parameter eikonal of $Q$-$\bar{Q}$ scattering was calculated. Clearly, Eq.~(\ref{Eq:12}) suggests that when any two of the three quarks which contribute to the singlet "nucleon" are well separated, there results a restoring potential $V(r_{ij})$ of the same form as that of Eq.~(\ref{Eq:7}).  There are, of course, corrections to this confining potential, those which are obvious from the approximations used above, as well as those corresponding to spin and angular momentum, which have been completely neglected.

\section{\label{SEC:SEC3}Estimation of the "Model Pion" Mass}

This Section contains an estimate of the effects of the above binding potential for the simplest case of the "pion", a ground state of the $Q$-$\bar{Q}$ system calculated using the simplest minimization technique~\cite{Quantics}, as well as a second minimization with respect to a ratio of terms introduced in this analysis.  In this simplest, non-relativistic estimation, the Hamiltonian of this system is given by
\begin{equation}
\mathcal{H} = 2m + \frac{1}{m} p^{2} + V(r),
\end{equation}

\noindent where $1/m$ denotes the inverse of the reduced mass of this equal-mass quark system.  The ground-state energy is then given by the replacement of $p$ by $1/r$, and the subsequent minimization of the eigenvalue of this Hamiltonian with respect to $r$.  Assuming $\xi \ll 1$, one obtains
\begin{equation}
\mu r = \left( \frac{2}{\xi} \, \frac{\mu}{m} \right)^{1/3},
\end{equation}

\noindent and
\begin{equation}\label{Eq:13}
E_{0} = 2m + \frac{3}{2} \xi \mu \left( \frac{2}{\xi} \, \frac{\mu}{m} \right)^{1/3},
\end{equation}

Here, $\mu$ and $\xi$ are the two parameters required by this analysis in order to have a non-zero binding potential; and while their existence has been introduced as a necessary assumption for the interacting $Q$-$\bar{Q}$ system, there is another unknown quantity, the quark mass $m$, which must play an important role.  If we take the position that $\mu$ and $\xi$ will appear as a result of a more fundamental QFT (in which the transverse momenta or position coordinates of the quanta of the QCD fields can never be measured with precision), then we are free to consider which values of $m$, or of the ratio $\mu / m$, can lead to the lowest value of $E$.  In so doing, we are asking if there might exist a dynamical reason for the relatively small value of the pion mass, in this approximation to the actual pion; that is, whether approximate chiral symmetry has a dynamical basis.

Following this approach, and using the small-$\xi$ limit of Eq.~(\ref{Eq:13}),
\begin{equation}\label{Eq:14}
E_{0} = \mu \left[ 2 \left(\frac{\mu}{m}\right)^{-1} + 3 \left(\frac{\xi}{2}\right)^{2/3} \, \left( \frac{\mu}{m} \right)^{1/3} \right],
\end{equation}

\noindent upon variation with respect to $x = \mu / m$, $E_{0}(x)$ will have an extremum at $x_{0} = 2^{3/4} \left( \frac{2}{\xi} \right)^{1/2}$; and since
\begin{equation}
\left. \frac{\partial^{2} E_{0}(x)}{\partial x^{2}} \right|_{x_{0}} = \mu \, 2^{-5/4} \, \left(\frac{\xi}{2}\right)^{3/2} \, \left( 2 - \frac{2}{3} \right) > 0,
\end{equation}

\noindent  that extremum is a minimum.  Substituting the value of $\mu / m = 2^{3/4} \, \left(\frac{2}{\xi}\right)^{-1/2}$ into Eq.~(\ref{Eq:14}) yields
\begin{equation}\label{Eq:15}
E_{0} = \mu \, \xi^{1/2} \, 2^{-1/4)} \, [1 + 3],
\end{equation}

\noindent from which one infers that there is three times as much energy in the gluon field and $Q$-$\bar{Q}$-kinetic energies as in the quark rest masses.  Intuitively, one expects that $E \sim m_{\pi} \sim \mu$, which suggests that $\xi \sim \sqrt{2}/16$; and, finally, one then has an estimate of the quark mass, in terms of $\xi$ and $\mu$.  The value of $m_{\pi}$ may be inferred from the fall-off of the (approximate) Yukawa nucleon-nucleon potential for moderate impact parameters, corresponding to separation distances in the "one-pion-exchange" region of that potential; and this will be discussed in a subsequent paper.

Of course, this double-minimization calculation of $E_{0}$ cannot be exactly identified with the precise pion mass, because of the approximations made above, as well as the omission of more complicated singlet terms, such as the contributions coming from $Q Q \bar{Q} \bar{Q}$ terms.  But the Physics seems to be reasonably correct.

A similar minimization analysis can be made for the nucleon ground state, but there then are other degrees of freedom which should be included, and treated properly by a serious attempt at a solution of such a three-body, relativistic problem.  This we must leave to others, whose numerical abilities far exceed our own.

\section{\label{SEC:Summary}Summary}

In this paper we have indicated how analytic binding potentials "naturally" appear in terms of the eikonal amplitudes considered, as the simplest, non-trivial examples of the new, gauge-invariant approach to analytic QCD.  In the necessary shift from "ideal" to "realistic" QCD, we have made use of a simple formalism~\cite{Fried2011_QCD3} which incorporates transverse imprecision in terms of a small, real, and positive parameter $\xi$, whose numerical value, and that of the ratio of $\mu/m$ can be inferred by a comparison with the measured pion mass.  Even though $\xi$ appears small, the contribution of the binding potential to the lowest bound-state-model-pion mass is large.  One can expect the same sort of relationship between the nucleon mass and the parameters $\xi$, $\mu$, and $m$ to occur for the $QQQ$ model of a nucleon.

	The next paper of this Series will deal with Hadronic Scattering Amplitudes, which employs the same set of parameters, and which will produce effective, Yukawa-type exponentially-decreasing scattering potentials between bound, singlet combinations of quarks and anti-quarks, scattering at distances larger than typical impact-parameter distances of the binding calculation of this paper; and the exponential fall-off of the scattering potentials will also depend on the same set of parameters.  To this end it will be necessary to enlarge the scope of the non-perturbative terms included by removing, in a simple, illustrative manner, the quenched approximation of the binding calculation above.  We emphasize that these calculations incorporate the exchange of all possible gluons between realistic quarks and anti-quarks in sums over all possible, relevant Feynman graphs: the functional techniques used here are intuitive, simple, and manifestly gauge-invariant.



\begin{acknowledgments}
One of us (HMF) would like to acknowledge that this publication was made possible through the support of a grant from the John Templeton Foundation. The opinions expressed in this publication are those of the authors and do not necessarily reflect the views of the John Templeton Foundation.
\end{acknowledgments}



\end{document}